\documentstyle[aps,epsf,twocolumn]{revtex}
\newcommand{\be}{\begin{eqnarray}}
\newcommand{\ee}{\end{eqnarray}}
\newcommand{\ben}{\begin{eqnarray*}}
\newcommand{\een}{\end{eqnarray*}}
\newcommand{\beq}{\begin{equation}}
\newcommand{\eeq}{\end{equation}}

\newcommand{\cdless}{\!\cdot\!}
\begin{document}
\preprint{SUNY-NTG-97-54}
\draft
\tighten
\title{\bf Photon Rates for Heavy-Ion Collisions
from Hidden Local Symmetry}

\author {Mikl\'{o}s-\'{A}d\'{a}m Hal\'{a}sz$^1$, James V. Steele$^2$,
Guo-qiang Li$^1$, and Gerald E. Brown$^1$ }
\address{$^1$Department of Physics and Astronomy, 
SUNY at
Stony Brook, Stony Brook, New York 11794-3800, USA\\
$^2$Physics Department, The Ohio State University, 174 W 18th Ave,
Columbus, Ohio 43210-1168, USA}

\date{\today}
\maketitle
\begin{abstract}

We study photon production from the hidden local symmetry approach
that includes $\pi$, $\rho$, and $a_1$ mesons and 
compute the corresponding photon emission rates from a hadronic
gas in thermal equilibrium.   
Together with experimental radiative decay widths of the background,
these rates are used in a relativistic  
transport model to calculate single photon spectra 
in heavy-ion collisions at SPS energies. 
We then employ this effective theory to test three scenarios
for the chiral phase transition in high-temperature nuclear matter 
including decreasing vector meson masses.
Although all calculations respect the upper bound set by the WA80
Collaboration, we find the scenarios could be distinguished with more
detailed data. 

\end{abstract}
\pacs{12.40-y, 25.75-q, 25.20Lj, 13.75Lb }

\narrowtext

\section{Introduction}

Experiments due to be performed over the next few decades are aimed at
achieving a quark-gluon plasma, thereby hopefully granting the ultimate
confirmation of QCD as the theory of strong interactions and also
settling the debate over the phenomenology of hadronic
matter at high density.
However, data currently available from experiments carried out at the
CERN SPS in the form of dilepton and photon spectra offer information
relevant to many outstanding issues, such as the enhancement of
low-mass dileptons in central heavy-ion collisions reported by the
CERES and HELIOS collaborations \cite{ceres,helios95},
anomalous $J/\psi$ suppression in central Pb+Pb collisions
by the NA50 collaboration \cite{na50}, and finally the limit on
single photon 
spectra from central heavy ion collisions set by the WA80
collaboration \cite{wa80}.

There has been intense theoretical activity in the past few years to
predict the behavior of hadronic matter at high density as chiral
symmetry restoration and deconfinement are approached. However, the
general physical picture of QCD phase transitions is still not agreed
upon. Lattice QCD lacks the techniques to surmount the
finite-density problem and other models differ in opinion to the
important effects.  In particular, there is nothing to prohibit chiral
symmetry from being restored via the Georgi vector limit
\cite{georgi}.  

In this limit, the $\rho$ becomes massless.  Its longitudinal
polarization must disappear, transforming into a scalar-isovector
particle identified as the chiral partner of the pion.  The $a_1$ is
the chiral partner of the 
$\rho$ and so also must become massless.  The
interesting question is the nature of the mass decrease of the vector
mesons and whether heavy ion collisions will be able to distinguish
between them. The decrease in general is an effect still in contention
although it is able to explain the dilepton spectrum from CERES within
a transport code calculation \cite{li95}.

Based on the restoration of scale invariance of QCD for low momentum scales,
Brown and Rho \cite{br91} suggested that the mass of non-strange vector mesons 
should decrease in dense matter, together with the chiral condensate.
There are a number of arguments that emerged that are in favor of Brown-Rho
scaling \cite{sumrule}. However, finite-temperature calculations based
on effective 
Lagrangians \cite{harada97,rapp} and some lattice results
\cite{gottlieb96} find that 
the $\rho$ mass does not vanish in the chiral limit. 
 
The aim of this paper is to compare the predictions of decreasing 
in-medium vector meson masses (whether relying on the Georgi vector
limit or not)
with the WA80 photon spectra. The most important processes are
$\pi \pi \rightarrow \rho \gamma$ and $\pi\rho \rightarrow \pi \gamma$
as well as the decay $\rho\rightarrow\pi\pi\gamma$ as determined in
Ref.~\cite{kapusta}. 
The $a_1$ resonance is also important for photon rates, as discussed in
\cite{xiong,song93,kim96}, and was treated incompletely 
in the transport calculations of Ref.~\cite{librown97}.
Furthermore, different scenarios for the density dependence of matter
such as the Georgi vector limit have not been addressed in the context
of photon rates in the past. We therefore use the extended hidden
local symmetry Lagrangian to make easy tests of different scenarios
through the adjustment of the parameters. The Georgi
vector limit is possible in the hidden gauge description as described
in the next section.  The massive Yang-Mills Lagrangian, as used in 
\cite{song93}, is to some extent a special case of this Lagrangian, 
corresponding to a specific parameter choice and a different gauge-fixing 
scheme. 

In general, the effective parameters -- masses and couplings -- may be
seen as a result of two distinct steps. First, the low-energy limit of QCD
defines the physics of hadrons \it in free space \rm due to chiral
symmetry breaking and confinement. Second, the physics of hadrons at
high 
density and temperature is obtained through many-body effects with the
above mentioned low-energy interaction. 

One might argue to what extent this dichotomy is a natural thing.
Whether it is
 suggested by the very existence of hadrons, or is a result of the historic
development of strong interaction physics.
There is no good reason to exclude 
medium effects from either of these steps. For instance, finite-temperature 
lattice QCD calculations, in principle complete, consider ensembles of a 
very small number of particles. Temperature effects at this level are mainly 
due to modification of the QCD vacuum, and may be 
compared to
internal structure modifications of hadrons, rather than many-body effects.
Traditional many-body calculations, on the other hand, use the same parameters
in their basic Lagrangian at all temperatures and densities.
Each of these approaches 
taken separately might miss  part of the picture. 

In this paper, we do not attempt to
predict medium modifications of the vector meson masses. Instead, we
assume all medium effects are induced by Brown-Rho scaling of the
vector meson masses and the pion decay constant through three
different scenarios as presented below.  We therefore can work both in
free space and at high temperature and density by adjusting the
couplings within the hidden gauge theory framework as will be shown in
the next section.

The remainder of the paper is structured as follows. In Section~II we
briefly review the hidden local symmetry (HLS) formalism extended to
both vector and axial-vector mesons and its fundamental results. Then
we propose a scheme of parameter assignment which leaves only the
vector meson masses as external parameters. In Section~III we present
our main 
analytical results, the $\rho \rightarrow \pi\pi\gamma$ decay width
and cross sections for $\pi\pi\rightarrow\rho\gamma$ as well as
$\pi\rho\rightarrow\pi\gamma$, and show the thermal equilibrium photon
rates derived from them. In Section~IV we discuss different ways to
implement Brown-Rho scaling and their effect on the results of the
previous section. In Section~V we describe the transport model we use and
then present the predictions for photon rates from S+Au collisions
with the different implementations of Brown-Rho scaling.  Finally,
Section~VI is devoted to a summary and conclusions.

\section{Hidden Local Symmetry Model}

We start with a brief review of the extended HLS model. The reader is
referred to \cite{hidden} for further details.
We would like to include the dynamics of both the $\rho$ and $a_1$
mesons and at the same time have a mechanism which allows for the
Georgi vector limit \cite{georgi}.  This has been done for the $\rho$
meson alone using an $SU(2)$ hidden local symmetry \cite{hidden}.  In that
case the Lagrangian depends on a single parameter $a$ with the KSRF
relation \cite{ksfr}  $g_\rho = 2f_\pi^2 g_{\rho\pi\pi}$ being a general 
consequence
of the theory.  Imposing $a=2$ gives the universality of the
$\rho$-couplings, a second KSRF relation $m_\rho^2=2g_{\rho\pi\pi}^2
f_\pi^2$, and $\rho$ dominance which are all phenomenologically
motivated.  The Georgi vector limit is approached as $a\to1$.
In this limit, the ordering of the hidden local symmetry Lagrangian in
terms of the number of derivatives can be vindicated since both the
$\pi$ and $\rho$ are light.  Away 
from the vector limit this ordering of terms can still be considered
valid. 

In order to include the $a_1$, we need to extend the local symmetry to
$SU(2)_L\times SU(2)_R$.  This increases the number of terms in the
most general Lagrangian to give
\be
{\cal L} = a\; {\cal L}_V + b\; {\cal L}_A + c\; {\cal L}_M + d\;
{\cal L}_\pi + {\cal L}_{\rm kin.}
\label{lagr}
\ee
with the kinetic terms for the vector $V_\mu$ and axial-vector $A_\mu$
fields included in the last term.  In addition we can add a global
symmetry $SU(2)\times SU(2)$ to describe external vector ${\cal
V}_\mu$ (including the photon) and axial-vector ${\cal A}_\mu$ fields.

Defining the conventional non-linear sigma field for the pion in terms
of three $SU(2)$ fields: $U(x)=\xi_L^\dagger(x) \xi_M(x) \xi_R(x)$, we
have \cite{hidden}
\ben
{\cal L}_{V,A} &=& \frac{f_\pi^2}4 {\rm Tr} \left| D_\mu \xi_L
\xi_L^\dagger  \pm \xi_M D_\mu \xi_R \xi_R^\dagger \xi_M^\dagger
\right|^2 
\\
{\cal L}_M &=& \frac{f_\pi^2}4 {\rm Tr} \left| D_\mu \xi_M
\xi_M^\dagger \right|^2
\\
{\cal L}_\pi &=& \frac{f_\pi^2}4 {\rm Tr} \left| D_\mu \xi_L
\xi_L^\dagger - \xi_M D_\mu \xi_R \xi_R^\dagger \xi_M^\dagger - D_\mu
\xi_M \xi_M^\dagger \right|^2
\\
D_\mu \xi_{R,L} &=& \partial_\mu \xi_{R,L} - i g(V\pm A)_\mu \xi_{R,L}
+ ie \xi_{R,L} ({\cal V}\pm {\cal A})_\mu
\\
D_\mu \xi_M &=& \partial_\mu \xi_M - ig (V-A)_\mu \xi_M + ie \xi_M
({\cal V} + {\cal A})_\mu .
\een
The choice $d=1-b+b^2/(b+c)$ ensures the pion kinetic term in
${\cal L}$ has unit coefficient.  Fixing the gauge $\xi_M=1$ and
$\xi_L^\dagger = \xi_R = \exp(i\pi/f_\pi)$, there is $\pi$-$a_1$
mixing present in ${\cal L}_A$ which can be eliminated by the shift
\[
A_\mu \to A_\mu + \frac{b}{b+c}\, \frac1{f_\pi g} \partial_\mu \pi .
\]
The gauge choice for $A$ and $V$ could introduce 
two other unphysical scalar fields ($\sigma$ and $p$) that parameterize 
the decomposition of $U(x)$ above. This also leads to mass generation for
these new fields through the usual
Stueckleberg construction. However, choosing the unitary gauge rids
the Lagrangian of the unphysical fields at merely the expense of making the
vector-field propagators non-transverse. Also, the vector part reduces to 
the CCWZ
effective Lagrangian in this gauge \cite{georgi}. We will stick to this
gauge throughout for convenience.

Unfortunately, upon adding the $a_1$, the first KSRF relation develops a
strong momentum dependence and the decay width
$\Gamma_{a_1\to\pi\gamma}$ vanishes. In addition, the width
$\Gamma_{a_1\to\rho\pi}$ is half of its experimental value. (For a
full discussion of these points see \cite{hidden}.) These
unattractive features can be rectified by including higher derivative
terms.  The added part is chosen to be \cite{hidden}
\ben
&&\delta {\cal L} = -{\cal L}_4 + {\cal L}_5 + {\cal L}_6 
\\
{\cal L}_4 &=& \frac{i}4 {\rm Tr} \left[ \alpha_M^\mu \alpha_M^\nu
F^{(L)}_{\mu\nu} + \xi_M^\dagger \alpha_M^\mu \alpha_M^\nu \xi_M
F^{(R)}_{\mu\nu} \right]
\\
\\
{\cal L}_5 &=& -\frac{i}4 {\rm Tr} \left[ \alpha_L^\mu \alpha_M^\nu
F^{(L)}_{\mu\nu} - \alpha_R^\mu \xi_M^\dagger \alpha_M^\nu \xi_M
F^{(R)}_{\mu\nu} \right] + {\rm h.c.}
\\
{\cal L}_6 &=& \frac{i}4 {\rm Tr} \left[ \xi_M \alpha_R^\mu
\xi_M^\dagger \alpha_M^\nu F^{(L)}_{\mu\nu} - \xi_M^\dagger
\alpha_L^\mu \alpha_M^\nu \xi_M F^{(R)}_{\mu\nu} \right] + {\rm h.c.}
\een
with $\alpha^\mu = (D^\mu \xi) \xi^\dagger$ and $F^{(R,L)}_{\mu\nu}$
defined as the field strength for the field combination $(V\pm
A)_\mu$.  For $a=2$ the first KSRF relation $g_\rho =
2f_\pi^2 g_{\rho\pi\pi}$ and vector dominance are regained.  The $a_1$
decays take 
on the reasonable values $\Gamma_{a_1\to\rho\pi}=360$ MeV and 
$\Gamma_{a_1\to\gamma\pi}=320$ keV.

Saturating the two Weinberg sum rules with the narrow width approximation
\cite{weinberg}
gives the two relations\footnote{Adding widths to the resonances
changes the quantitative result for the parameter $a$ by less than a
percent.} 
\be
g_\rho=g_{a_1}\qquad\qquad m_{a_1}^2 = \frac{a}{a-1} m_\rho^2 \equiv
\frac{m_\rho^2}{r} 
\label{mass}
\ee
which are satisfied for $b=a$ and $c=a/(a-1)$ in eq.~(\ref{lagr}).
This leaves only the one 
parameter $a$ to govern the evolution of the Lagrangian towards the
Georgi vector limit.  For $a=2$, the familiar relation $m_{a_1}^2 =
2m_\rho^2$ is reproduced.  This gives $m_{a_1}=1090$ MeV whereas the
experimental value is $m_{a_1}\simeq 1230$ MeV.  This discrepancy suggests
that $a=2$ is not the proper value in free space and it should be 
closer to $a=1.64$.  

Setting $a=1.64$ changes the other physical observables as well.
First of all, the $\gamma\pi\pi$ vertex is not
entirely vector dominated,
resulting in a non-zero $g_{\gamma\pi\pi}=0.18e$ direct coupling to
the photon.  (The total contribution including the $\rho$-$\gamma$
mixing is still, of course, equal to $e$.)  The $\rho$-coupling
universality is modified to become $g_{\rho\pi\pi}=0.96 g$, still
within reason. 
In fact, the two KSRF relations can be written for
general $a$ as
\[
g_{\rho\pi\pi} = \frac12 ag(1+\xi_r)
\qquad\qquad
m_\rho^2 = \frac{4}{a(1+\xi_r)^2} g_{\rho\pi\pi}^2 f_\pi^2
\]
with $\xi_r = 2r(1-2r)$ and $r$ defined in eq.~(\ref{mass}).  Using
$f_\pi=92.4$ MeV and $g_{\rho\pi\pi}^2/4\pi\simeq 3.0$ \cite{pdata}
improves the second KSRF relation prediction for the $\rho$ mass from
$m_\rho=800$ MeV ($a=2$) to $760$ MeV ($a=1.64$). Notice that Georgi's
relation $m_\rho^2=4g_{\rho\pi\pi}^2 f_\pi^2$ is reproduced for
$a=1$. The pion vector
radius for $a=2$, $\langle r^2\rangle_V^\pi = 0.39$ fm$^2$, is
slightly below the data $\langle r^2\rangle_V^\pi = 0.44\pm0.03$
fm$^2$.  For general $a$ the value is 
\[
\langle r^2\rangle_V^\pi = \frac{3(1+\xi_r)}{g^2f_\pi^2} 
\]
which for $a=1.64$ results in $0.46$ fm$^2$ in good agreement with the
data.  The pion polarizability can then be determined from the
Das-Mathur-Okubo relation \cite{das} and gives values consistent with 
data for both
values of $a$.

The decays of the $a_1$ meson also are modified.  The total width is
dominated by $\Gamma_{a_1\to\rho\pi}$ which changes from $360$ MeV to
$430$ MeV with $a=1.64$, still about equal to the experimental value
of $400$ MeV.  The surprise is the radiative decay 
$\Gamma_{a_1\to\gamma\pi}$.  It changes
from $320$ keV to $50$ keV as given by the formula
\[
\Gamma_{a_1\to\gamma\pi}= \left(\frac{e}{g} \right)^2 
\frac{(3r-1)^2|{\bf p}|^3}{12\pi f_\pi^2} .
\]
This is very different from the experimental value $630$~keV also
obtained 
by vector meson dominance.  This is due to the contribution and
interference from the graph with a direct photon.  This
is responsible for pulling the width down to $320$ keV with
$a=2$ ($r=1/2$) and almost vanishing for $a=1.64$ ($r\simeq 1/3$).  

Therefore, not assuming total vector dominance in nature actually
improves most quantities towards their measured values by about
$10\%$ with the exception of the decay width $\Gamma_{a_1\to\gamma\pi}$ 
which nearly vanishes to ${\cal O}(e^2)$.

\section{Analytical results}

\begin{figure}
\begin{center}
\leavevmode
\epsfxsize=3in
\epsffile{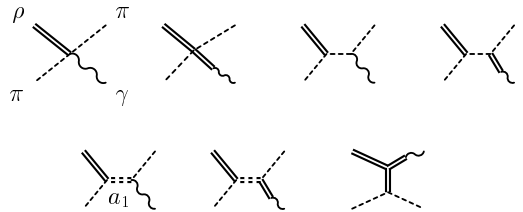}
\end{center}
\caption{\label{fig1} The general diagrams for the matrix element
involving $\rho\pi\pi\gamma$.}
\end{figure}

Although we will also consider the effects of the $\omega$, $\eta$,
and $\eta^\prime$ mesons, these particles are long enough lived to 
mostly decay after freeze-out and therefore we can take the 
experimental values for their partial widths.  The only decay 
abundant enough to consider is then $\rho\to\pi\pi\gamma$.  
For the photon production from two-body collisions, only $\pi\pi\to
\rho\gamma$ and $\pi\rho\to \pi\gamma$ are important enough to
consider for the temperature region relevant to heavy-ion
collisions at SPS energies \cite{kapusta}. These three processes 
are all related to the same matrix element quoted in Appendix A.  
Electromagnetic gauge invariance can easily be checked on our expressions. 
The contributing graphs are shown in Fig.~\ref{fig1}.

%%%%%%%%%%%%%%%%%%%%%%%%%%%%%%%%%%%%%%%%%%%%%%%%%%%%%%%%%%
%%%%%%%%%%%%%%%%%%%%%%%%%%%%%%%%%%%%%%%%%%%%%%%%%%%%%%%%%%%%%%

First focusing on the decay $\rho\to\pi\pi\gamma$, the result 
in free space for $a=2$ $(1.64)$ is  
\[
\Gamma(\rho^0\to\pi^+\pi^-\gamma)=1.6 \,\,(1.8) \,\, \hbox{MeV}
\]
Both values are in reasonable agreement with the experimental data 
$1.5\pm0.3$ MeV \cite{pdata}. In the same model, we can calculate the
radiative decay width of charged rho mesons, and we obtain 
$\Gamma (\rho\to\pi\pi\gamma)=0.88$ $(0.98)$ MeV after
isospin averaging, showing the $\rho^\pm$ together give slightly less
contribution than the $\rho^0$ alone.

\begin{figure}
 \begin{center}
  \leavevmode
  \hbox{
   \hspace{-.5in}
\vspace{-0.5in}
  \epsfxsize=3.5in
   \epsffile{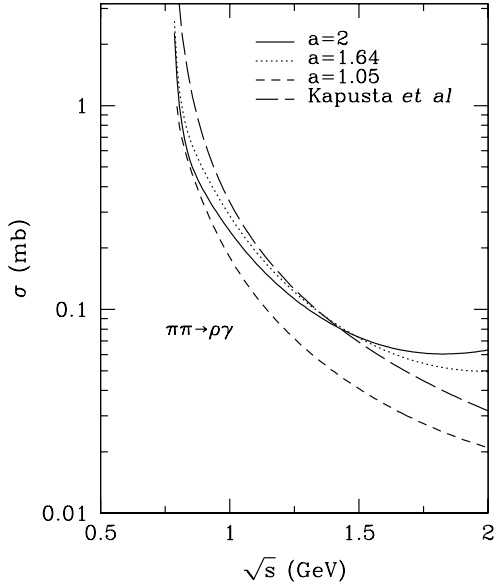}
  }
 \end{center}
\vspace{-0.2in}
 \caption{
   \label{fig2} 
   The total cross section for
   $\pi\pi\to\rho\gamma$ for $a=2$, $1.64$ and $1.05$. 
The value of $g$ is determined by $m_{\rho}^2=a g^2 f_{\pi}^2$. 
The result of Ref.~\protect\cite{kapusta} is shown for comparison.
 }
\end{figure}

\begin{figure}
 \begin{center}
  \leavevmode
  \hbox{
   \hspace{-.5in}
\vspace{-0.5in}
   \epsfxsize=3.5in
   \epsffile{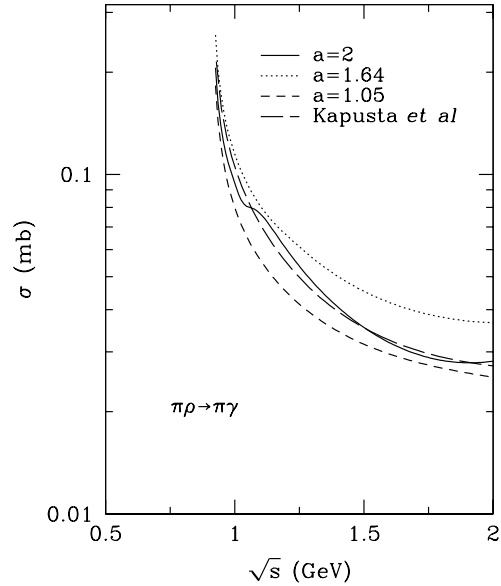}
  }
 \end{center}
\vspace{-0.2in}
 \caption{
   \label{fig3} 
   The total cross section for
    $\pi\rho\to\pi\gamma$
    for $a=2$, $1.64$ and $1.05$. 
The value of $g$ is determined by $m_{\rho}^2=a g^2 f_{\pi}^2$. 
The result of Ref.~\protect\cite{kapusta} is shown for comparison.
 }
\end{figure}

The cross sections for $\pi\pi\to\rho\gamma$ 
and $\pi\rho\to\pi\gamma$ can be calculated in the same
way, and the results are shown in Fig.~\ref{fig2} and \ref{fig3} for
three values of $a$. A momentum dependent width for the $a_1$ mesons
as calculated in the HLS Lagrangian was included.  
It is useful to compare our results with those of Kapusta,
Lichard, and Seibert \cite{kapusta}, which did not include the effects
of the $a_1$ meson.  

%%%%%%%%%%%%%%%%%%%%%%%%%%%%%%%%%%%%%%%%%%%%%%%%%%%%%%%%%%
%%%%%%%%%%%%%%%%%%%%%%%%%%%%%%%%%%%%%%%%%%%%%%%%%%%%%%%%%%%%%%

For $\pi\pi\to\rho\gamma$, the $a=2$ and $a=1.64$ results are similar
and near the results of Kapusta, {\em et al.}, within the pertinent
values of $\sqrt{s}$.   Likewise, the $\pi\rho\to\pi\gamma$
result for $a=2$ (which has a visible bump close to $\sqrt{2} m_\rho$) 
drops quickly to the Kapusta, {\em et al.}, cross
section due to 
the broadening in the momentum dependent $a_1$ width.  The $a=1.64$
result however, is larger by almost a factor of two, showing there is
no simple connection between having a small $\Gamma_{a_1\to\gamma\pi}$ width
and the cross section in the $\pi\rho$ channel. Therefore we will use
$a=1.64$ for the free space calculations below and not concern
ourselves further with the $a=2$ case.   As a
consistency check, our results reduce essentially to those of Kapusta, 
{\em et al.},
when we turn off the $a_1$ effects, as they should.

The $a=1.05$ results lead to substantially smaller cross sections,
indicating that less photons will be produced from these processes
when chiral symmetry restoration is approached.  This is especially
seen when comparing to the free space result ($a=1.64$).  Including
the scaling of the parameter $g$ as well will lead to a decrease of
the vector meson masses as discussed in the next section.  The
suppression of cross sections seen here will be counterbalanced by an
enhancement from the increase in phase space and the rates will not
appreciably change.

%%%%%%%%%%%%%%%%%%%%%%%%%%%%%%%%%%%%%%%%%%%%%%%%%%%%%%%%%%
%%%%%%%%%%%%%%%%%%%%%%%%%%%%%%%%%%%%%%%%%%%%%%%%%%%%%%%%%%%%%%

Emission rates for photons from a gas of hadrons at thermal equilibrium
can be readily computed from the matrix element for a given process.
For the process $1+2\to 3+\gamma$, the differential rate is given by
\ben
&&E_\gamma \frac{dR}{d^3p_\gamma} = \frac{\cal N}{2(2\pi)^3} \int 
\frac{d^3 p_1}{(2 \pi)^3 2 E_1}
\frac{d^3 p_2}{(2 \pi)^3 2 E_2}
\frac{d^3 p_3}{(2 \pi)^3 2 E_3}
\left| {\cal M} \right|^2
\\
&&\qquad(2 \pi)^4 \delta^4(p_1+p_2-p_3-p_\gamma)
f(E_1) f(E_2) \left(1 + f(E_3)\right)
\een
where ${\cal M}$ is the scattering amplitude,
$f(x)=(\exp(x/T)-1)^{-1}$ is the 
Bose-Einstein distribution, and ${\cal N}$ is the degeneracy factor
which equals the number of distinct incoming states considered
in the matrix element. 
Since the matrix element depends only on the 
Mandelstam variables $s=(p_1 + p_2)^2$ and $t=(p_1 - p_\gamma)^2$, we follow
\cite{kapusta} and insert integrals over each of these variables with delta 
functions ensuring the above definitions. The differential rate can then be 
expressed in terms of an exact quadruple integral as shown in
Appendix B. 

\begin{figure}
\begin{center}
\leavevmode
\hbox{
\hspace{-.5in}
\vspace{-0.5in}
\epsfxsize=3.5in
\epsffile{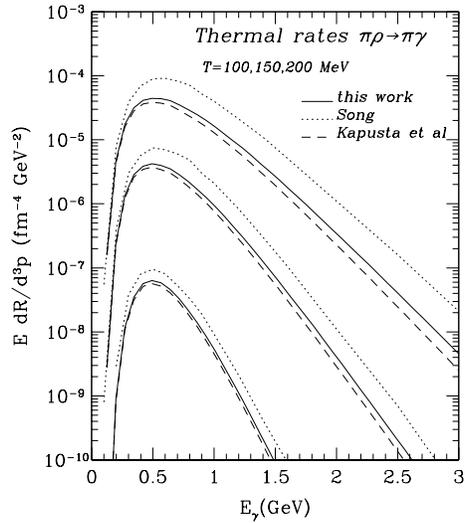}
}
\end{center}
\vspace{-0.2in}
\caption{ \label{fig4}
Thermal rates for the process $\pi\rho\to\pi\gamma$ 
as compared to those of Song 
and Kapusta  \it et al.\rm}
\end{figure}

\begin{figure}
\begin{center}
\leavevmode
\hbox{
\hspace{-0.5in}
\vspace{-.5in}
\epsfxsize=3.5in
\epsffile{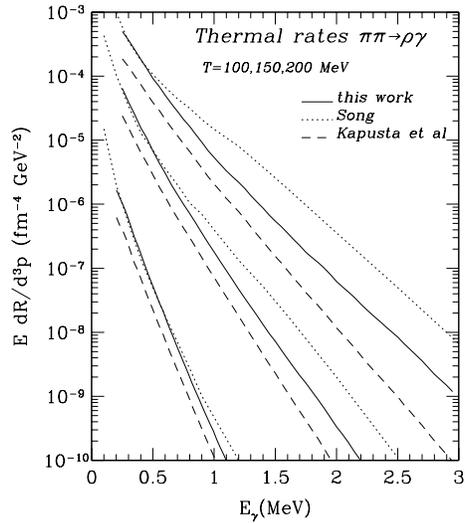}
}
\end{center}
\vspace{-0.2in}
\caption{ \label{fig5}
Same as Fig.\ref{fig4}, for
$\pi\pi\to\rho\gamma$.}
\end{figure}

\begin{figure}
\begin{center}
\leavevmode
\hbox{
\hspace{-0.5in}
\vspace{-0.5in}
\epsfxsize=3.5in
\epsffile{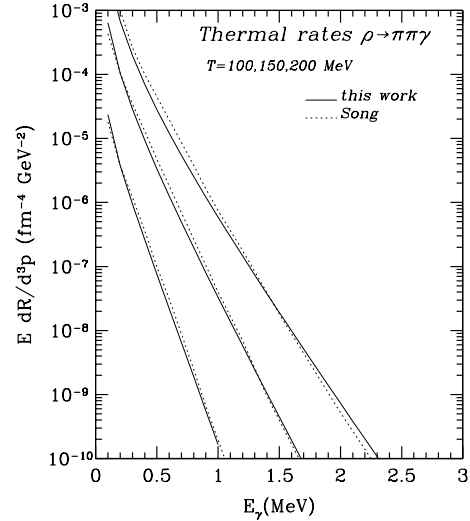}
}
\end{center}
\vspace{-0.2in}
\caption{ \label{fig6}
Same as Fig.\ref{fig4}, for 
$\rho\to\pi\pi\gamma$.}
\end{figure}

For the thermal emission rate from the decay $\rho\to\pi\pi\gamma$,
we used the approximate expression given in \cite{kapusta}.
As it is seen from Figs.~\ref{fig4}, \ref{fig5}, and \ref{fig6}, 
our thermal rates
for $a=1.64$ are considerably different from those of Song
\cite{song93} who included the $a_1$ meson within the massive
Yang-Mills approach but did not include a
momentum dependent width.  
For the $\pi\pi\to\rho\gamma$ process, the rates are similar to Song for
small photon energy but soon fall below. Although the massive Yang-Mills
Lagrangian is a special case of HLS, we have very different parameter
choices.  

%%%%%%%%%%%%%%%%%%%%%%%%%%%%%%%%%%%%%%%%%%%%%%%%%%%%%%%%%%%%%%
%%%%%%%%%%%%%%%%%%%%%%%%%%%%%%%%%%%%%%%%%%%%%%%%%%%%%%%%%%%%%%

%%%%%%%%%%%%%%%%%%%%%%%%%%%%%%%%%%%%%%%%%%%%%%%%%%%%%%%%%%%%%%
%%%%%%%%%%%%%%%%%%%%%%%%%%%%%%%%%%%%%%%%%%%%%%%%%%%%%%%%%%%%%%

The variable $a_1$ width is an important effect, leading to a very
broad resonance in the $s$-channel $\pi\rho\to\pi\gamma$ cross
section.  There the contribution of the $a_1$ is practically zero
for $\sqrt{s}\ge1.4$ GeV.  However, this does doesn't account for the
factor of approximately $3$ between our result for the thermal
emission rates and those of Song.  For the $\rho$ decay, our
result is almost identical to Song's, as expected from
our agreement on the corresponding decay width.  
At thermal equilibrium, the effects from this
decay are outshined by those due to the two-body processes
by almost two orders of magnitude. 

%%%%%%%%%%%%%%%%%%%%%%%%%%%%%%%%%%%%%%%%%%%%%%%%%%%%%%%%%%%%%%
%%%%%%%%%%%%%%%%%%%%%%%%%%%%%%%%%%%%%%%%%%%%%%%%%%%%%%%%%%%%%%

\section{Scenarios for the chiral phase transition}

The most attractive feature of HLS is that it describes fairly well
the dynamics of pions and vector mesons with only three parameters:
the pion decay constant $f_\pi$, the universal coupling constant $g$,
and the parameter $a$. From the construction of the HLS Lagrangian,
the latter two define the masses of the vector mesons
\be
m_\rho^2 = a g^2 f_\pi^2 \qquad\qquad m_\rho^2/m_{a_1}^2 = (a-1)/a.
\label{masses}
\ee
In practice, this relation may be reversed in order to use the vector 
meson masses to set the values of $a$ and $g$.

We set out to use this description for the physics of mesons at high 
temperature and baryon density. Our basic assumption is that, at least
for moderate conditions, the HLS description still
holds, only with changed parameter values. In other words, the effects of
temperature and medium may be described in terms of changing
$f_\pi$, $m_\rho$, and $m_{a_1}$ only. 

Undoubtedly, if our description is valid, renormalization effects as well
as temperature and medium effects within the HLS description \cite{harada97}
play a major
part in the evolution of our three parameters. On the other hand, there
may be more fundamental effects, which 
can not be predicted from within the effective theory. 
After all, spontaneous chiral symmetry breaking is a property of full QCD,
and the mechanism controlling it may or may not be captured in an approximate
description like ours.
 Therefore, we do not address within the HLS approach the  issue of 
how the three quantities evolve towards the phase transition.
 Instead, we will consider a number of scenarios based on very general 
arguments. 

The $\rho$ meson mass has been suggested as a possible order parameter for
the chiral phase transition \cite{AdamiBrown92}. If the phase transition 
is of second
order, then $m_\rho$ will smoothly decrease towards zero. According to the
scaling argument by Brown and Rho \cite{br91}, all other quantities 
such as $f_\pi$ and $m_{a_1}$ will be 
driven by the ratio $m_\rho^* / m_\rho$ ($m_\rho^*$ is the in-medium $\rho$ 
mass). 
Our aim is to find out the effect of such a picture on photon emission from 
heavy ion collisions. 

\it First scenario: \rm
The conventional wisdom would be to allow the $a_1$ mass to drop
towards zero along with the $\rho$ mass
\ben
\frac{m_\rho^*}{m_\rho} = \frac{m_{a_1}^*}{m_{a_1}}.
\een
However, this simple scaling
will freeze the HLS $a$ parameter at its free-space value as
can be seen from eq.~(\ref{masses}). This implies $a$ will never
attain its renormalization group fixed point of the HLS theory \cite{hidden},
 $a=1$,
which would lead to the Georgi vector limit\footnote{A one-loop
calculation of the renormalization group equations including the $a_1$
meson has not yet been done, but even a different fixed point $a=a_0$
will not change the final situations described in this paper.}.
The dependence of $g$ on the in-medium $\rho$ mass is determined by
the scaling dimension of $f_\pi$ as given by the
Gell-Mann-Oakes-Renner relation: 
\ben
\frac{m_\rho^*}{m_\rho} = \left(\frac{\langle\bar{\psi}\psi\rangle^*}
{\langle\bar{\psi}\psi\rangle^{\ }}\right)^{\!\!\alpha};
\left(\frac{f_\pi^*}{f_\pi}\right)^2 \!\!\! =
\frac{\langle\bar{\psi}\psi\rangle^*}
{\langle\bar{\psi}\psi\rangle^{\ }}
\;\Rightarrow\; \frac{f_\pi^*}{f_\pi} = 
\left(\frac{m_\rho^*}{m_\rho}\right)^{\!\!1/2\alpha} 
\een
The suggested values for $\alpha$ are $1/3$ \cite{br91}, $1/2$ 
\cite{Brown95}, and $1$ \cite{BrownBuballa96}. 
Since $g$ scales like $\langle\bar{\psi}\psi\rangle^{\alpha-1/2}$ from
the above and eq. \ref{masses},
choosing $\alpha=1/3$ would lead to an infinitely strong coupling in the
chiral limit contrary to the asymptotic freedom of QCD.  The choice of
$\alpha=1/2$ is therefore a limiting case, leading to
\ben
\frac{f_\pi^*}{f_\pi} = \frac{m_\rho^*}{m_\rho}
\een
and thus $g$ is frozen to its free-space value just like $a$ is.
This prescription is equivalent to the one used in \cite{librown97}.

\it Second scenario: \rm
As opposed to the complete disregard for the renormalization group
behavior of $a$ and $g$, we now take the fixed points to be
realized. If $a$ approaches $1$ as the $\rho$ mass goes to zero, the
ratio $m_{a_1}/m_\rho$ must diverge, as it can be seen  
from (\ref{masses}). This does not imply the $a_1$ mass can not
vanish, only that it must vanish slower than $m_\rho$.
As a limiting case, however, we take $m_{a_1}$ to be 
constant\footnote{In our simulations, we expect only relatively small effects
from partial chiral symmetry restoration. If the $m_{a_1}$  decreases
significantly slower than $m_\rho$, we may take it constant in
the first approximation.} and this determines $a$.  
Note that now the flow of $a$ and $g$ to their fixed
point values of $1$ and $0$ respectively are intimately connected such
that they 
conspire to give a constant $m_{a_1}$ as seen in eq.~(\ref{masses}).
Of the possible values cited, only $\alpha=1$ for the Brown-Rho
scaling drives $g$ to $0$. In conclusion, we take 
\ben
\left(\frac{f_\pi^*}{f_\pi}\right)^2 = \frac{m_\rho^*}{m_\rho} .
\een
This is one way of ``implementing'' the Georgi vector limit, without having
to address the issue of the fate of the $a_1$. If the $a_1$ becomes
massless along with the $\rho$, then its longitudinal polarization will
be a massless scalar, degenerate with the pion.

\it Third scenario: \rm A different physical picture that may be considered
\cite{hung96} as an alternative to the vanishing of $m_\rho$ in the
chiral limit is  based on lattice results indicating that $m_\rho$ 
and $m_{a_1}$ decrease with temperature but become equal at  
some finite value, not running to zero \cite{gottlieb96}. If a similar 
situation occurs with density, we may parameterize it by taking the
vector meson masses $m_\rho$ and $m_{a_1}$ as the sum of an invariant
piece $m_{deg}$, which is the same in both, and 
a piece that scales with the chiral order parameter to some positive power 
$\alpha$:
\ben
\phi\equiv\frac{M^*-m_{deg}}{M-m_{deg}} =
\left(\frac{\langle\bar{\psi}\psi\rangle^*} 
{\langle\bar{\psi}\psi\rangle^{\ }}\right)^{\!\!\alpha}
\een
for $M=m_\rho$ and $m_{a_1}$.  It is also possible to consider different
values of $\alpha$ for the two vector mesons, but for simplicity we
take them equal. Again, $f_\pi$ will scale with $\phi^{1/2\alpha}$ and
so in the limit $\phi\to 0$, $g$ is proportional to 
$\phi^{(\alpha-1)/2 \alpha}$, which
imposes the choice $\alpha=1$ among the ones mentioned above.
Therefore $f_\pi$ scales with $\phi^{1/2}$ and the matrix elements
sharply increase as $\phi$ approaches $0$.  Although there are only a
few occurrences of very low $m_\rho^*$, they are significantly enhanced
by this effect. In the following we take $m_{deg}=0.2 m_\rho$.

\begin{figure}
\vspace{-1.0cm}
\begin{center}
\leavevmode
\hbox{
\hspace{-0.5cm}
\epsfxsize=6.5in
\epsffile{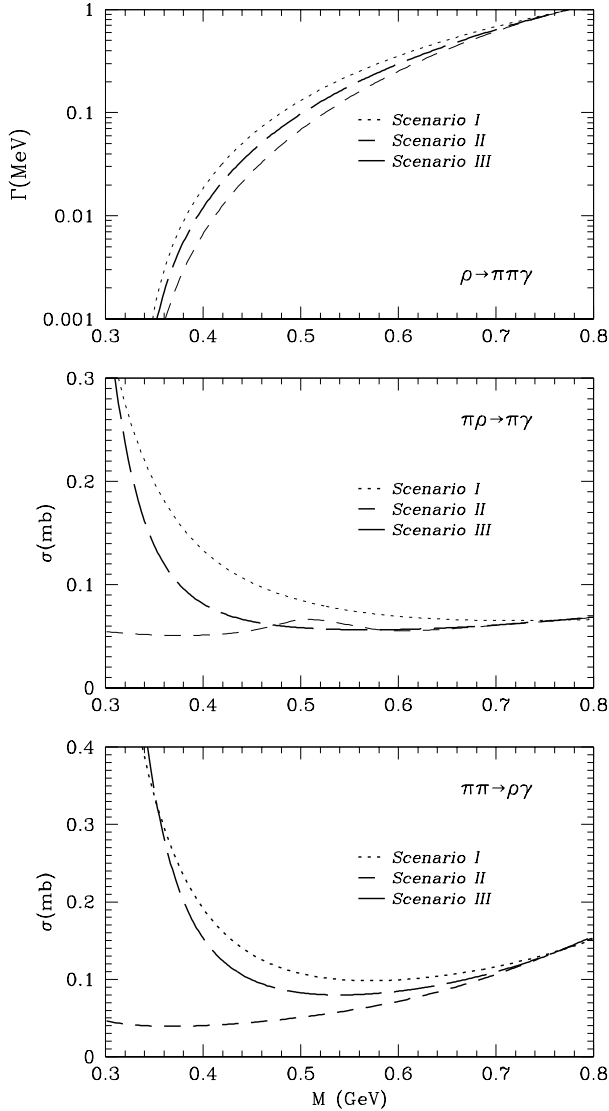}
}
\end{center}
\vspace{-0.5cm}
\caption{\label{fig7} Photon production cross sections at
$\protect\sqrt{s}=1200$ MeV and $\rho$ decay rate as a function of
$\rho$ mass, for the three different scenarios.} 
\end{figure}

\begin{figure}
\vspace{-1.3cm}
\begin{center}
\leavevmode
\hbox{
\hspace{0.3cm}
\epsfxsize=5.0in
\epsffile{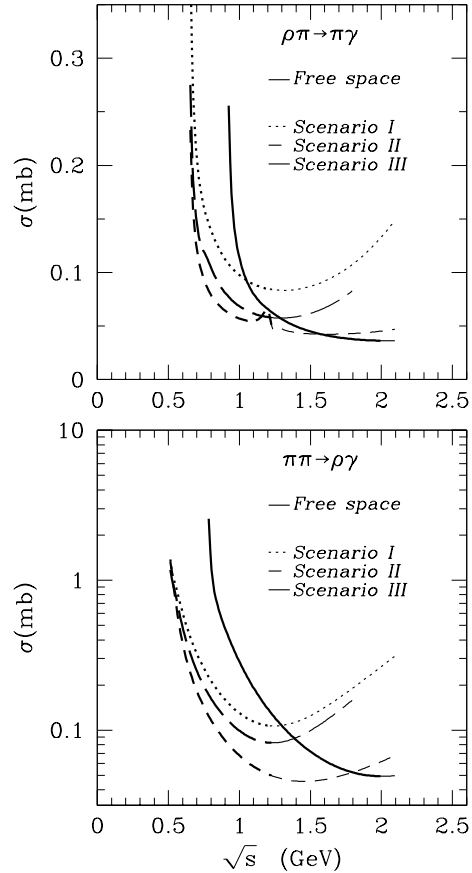}
}
\end{center}
\vspace{-0.5cm}
\caption{\label{fig8} Photon production cross sections
in free space and for a $\rho$ mass decreased to 500 MeV
in the three different scenarios.}
\end{figure}

Each of the three scenarios gives distinctly different results as
illustrated in Figs.~\ref{fig7} and \ref{fig8}. Fig.~\ref{fig7} shows
the dependence of the photon production cross sections (for fixed
$\sqrt{s}=1200$ MeV ) and $\rho$ radiative decay width
on the $\rho$ mass.  Since both the strict Brown-Rho scaling of Scenario I
and the lattice based scaling of Scenario III allow both vector meson
masses 
to drop together, we see an increase in the cross sections as the
$\rho$ mass decreases.  From this point of view, the main difference
in the three scenarios is the way in which the $a_1$ mass decreases:
either at the same rate as the $\rho$, not at all, or faster than the
$\rho$. This leads to a different rate of enhancement for lower $\rho$
masses, and no considerable change for the extreme case of the
Georgi-vector limit in Scenario II. This can
also be seen in Fig.~\ref{fig8} where the $\rho$ mass is fixed at
$500$ MeV.  The $a_1$ mass reduction in Scenario I
and III is accompanied by an increase in the cross sections.  

%%%%%%%%%%%%%%%%%%%%%%%%%%%%%%%%%%%%%%%%%%%%%%%%%%%%%%%%%%
%%%%%%%%%%%%%%%%%%%%%%%%%%%%%%%%%%%%%%%%%%%%%%%%%%%%%%%%%%%%%%

Notice that in Fig.~\ref{fig8} the cross sections for reduced $\rho$
mass tend to increase with $\sqrt{s}$ starting at approximately $1300$ MeV.
The same behavior is present in the free-space result, only at higher
values of $s$. This is certainly unphysical and comes from the terms
with high 
powers of momenta in the extended HLS Lagrangian introduced by the
redefinition of the $A_\mu$ field.
We are, after all, working with an effective theory which 
breaks down at high energies. In a more careful analysis, one should 
interpolate between the effective theory and asymptotically free QCD. 
For our purposes, we calculate our cross sections up to where they
start to increase again ($\sqrt{s}=2$ GeV for the free space case) and
then fix the cross section to that final value for larger 
values of $\sqrt{s}$. 
In Fig.\ref{fig8}, we indicate this by the
thicker lines below the cutoff $\sqrt{s}$. As long as the medium
effects we are considering are relatively small, the cutoff value stays
reasonably high, and our results should not be strongly influenced. However,
one should have this in mind when looking at the high $p_t$ region.

%%%%%%%%%%%%%%%%%%%%%%%%%%%%%%%%%%%%%%%%%%%%%%%%%%%%%%%%%%
%%%%%%%%%%%%%%%%%%%%%%%%%%%%%%%%%%%%%%%%%%%%%%%%%%%%%%%%%%%%%%

Naturally, the thermal equilibrium rates shown at the end of Section III
will increase dramatically for all three  scenarios as the vector
masses decrease. This is mostly due to the Boltzmann factor.
For example, if the $\rho$-meson mass is reduced to 500 MeV at $T=150$ MeV,
the thermal emission rate $\rho\pi\to\pi\gamma$ increases by about a
factor of 6. However, as we will discuss in the next section, this
effect is not the only factor to consider when applying these results to
heavy ion collisions.  In addition, the total pion multiplicity is
constrained by experiment.  Whereas the thermal 
rate calculation corresponds to a grand-canonical ensemble of mesons
which can produce excess pions at will, the real situation
in heavy-ion collisions is closer to an ensemble with fixed pion
number constrained by the hadronic observables.

\section{Photon Spectra in Heavy-Ion Collisions}

In studying medium effects in heavy-ion collisions, relativistic 
transport calculations \cite{koli96,kkl97} based on the 
Walecka-type model has been quite useful, providing
a thermodynamically consistent 
description of the medium effects through the scalar and vector
fields. In heavy-ion collisions at CERN-SPS energies, 
many hadrons are produced in the initial nucleon-nucleon 
interactions. This is usually modeled by the fragmentation of 
strings, which are the chromoelectric flux-tubes 
excited from the interacting quarks. One successful model 
for taking into account this non-equilibrium dynamics is the RQMD model 
\cite{sorge89}. To extend the relativistic transport model to heavy-ion 
collisions at these energies, we have used as initial conditions the 
hadron abundance and distributions obtained from the string fragmentation 
in RQMD.  

Further interactions and decays of these `primary' hadrons are then 
taken into account through a conventional relativistic transport model.
We include non-strange baryons with masses below 1.72 GeV, as well
as $\Lambda$, $\Lambda (1405)$, $\Sigma$, and $\Sigma (1385)$.
For mesons we include $\pi$, $\eta$, $\rho$, $\omega$, $\eta^\prime$,
$a_1$, and $\phi$, as well as $K$ and $K^*(892)$. Baryons are propagated in
their mean fields, which are assumed to be the same for all
non-strange baryons. The mean fields for hyperons are assumed 
to be 2/3 of that for non-strange baryons, based on the simple
quark counting rule.

In addition to propagation in mean fields, hadrons also 
interact under stochastic two-body collisions.
For baryon-baryon interactions, we include both elastic
and inelastic scattering for the nucleon, $\Delta (1232)$,
$N(1440)$, and $N(1535)$. Their cross sections are either
taken from Refs. \cite{arndt82,wolf93} or obtained using
the detailed balance procedure \cite{bert91}. The meson-baryon
interactions are modeled by baryon resonance formation and decay.
For example, the interaction of a pion with a nucleon
proceeds through the formation and decay of any of the $N$ or $\Delta$ 
resonances from the $\Delta (1232)$ up to the $N(1720)$. The formation
cross sections 
are taken to be of the relativistic Breit-Wigner form.
The meson-meson interactions are either formulated
through resonance formation and decay when the intermediate 
meson is explicitly included in our model, such as the $a_1$ meson,
or treated as a direct elastic scattering with a cross section
estimated from various theoretical models \cite{li95}. 

Photon creation is taken into account during the evolution of the transport
code through decays or two body processes. 
The experimental data for $\omega$ and $\eta'$ decays are used as
these particles are long-lived and mostly decay after escaping into
free-space. Photons can also be produced from the decay of baryon
resonances. These contributions are usually neglected
in hydrodynamical calculations \cite{sinha94,dumi95}, but are included
here through experimentally measured radiative decay widths  
\cite{pdata}. 

Otherwise, we include the three main contributors to photon
production: $\rho\to\pi\pi\gamma$, $\pi\pi\to\rho\gamma$, and
$\pi\rho\to\pi\gamma$ as calculated in the previous sections and
modify the two fundamental parameters $a$ and $g$ of the HLS
Lagrangian according to each of the three scenarios mentioned above.

\begin{figure}
\begin{center}
\hspace{-0.5in}
\epsfxsize=4.0in
\epsffile{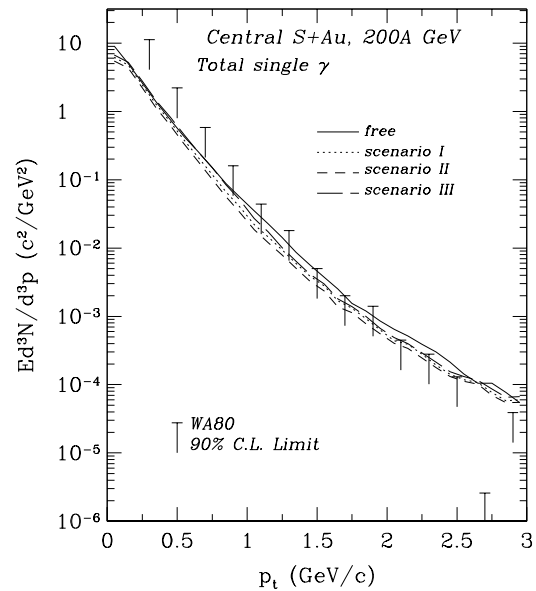}
\end{center}
\vspace{-0.3in}
\caption{\label{fig9} Total thermal photon spectra in
central S+Au collisions at 200 AGeV for the different scenarios
discussed.}
\end{figure}

In Fig.~\ref{fig9} we compare our thermal photon spectra
with the upper bound of the WA80 collaboration. 
Overall the results are below the upper bound for photons 
with transverse momenta well below 1 GeV. For higher transverse 
momenta, our results touch the upper bound of the experimental data. 
These observations are essentially the same as those of
Ref.~\cite{librown97}, although the contribution from two-body
collisions  
at high transverse momenta now becomes comparable to that of 
meson decay, due to a more realistic treatment of $a_1$ effects. 
This shows that whether the vector meson masses are reduced or not,
the rates do not change much since the opening up of phase space is
balanced against a decrease in the coupling constants.

%%%%%%%%%%%%%%%%%%%%%%%%%%%%%%%%%%%%%%%%%%%%%%%%%%%%%%%%%%
\vspace{-.5in}
%%%%%%%%%%%%%%%%%%%%%%%%%%%%%%%%%%%%%%%%%%%%%%%%%%%%%%%%%%%%%%

The fact that there is no dramatic increase in photon emission rates
would seem to contradict the naive expectations of thermal rate
calculations. 
Regardless of whether the dropping mass scenario is invoked or not,
any photon 
spectrum calculation must at the same time fit the observable pion
multiplicity. Through the use of a complete transport calculation as
described above, this quantity can be consistently obtained. 
As shown in Ref.~\cite{li95}, this could be achieved in a thermal model
by either decreasing the vector meson mass or increasing the pion chemical
potential by hand. Thus, for a meaningful comparison with the dropping mass
scenarios, one would need to use a large chemical potential in the bare mass
scenario, which would push up the thermal rates.
In this paper however, we 
do not use thermal equilibrium initial abundances. Instead, we evolve
the system from the same initial RQMD output for all cases considered.
This procedure guarantees the correct final pion yield without
additional assumptions such as a pion chemical potential \cite{lkbs}.

With a dropping $\rho$ mass, the $\rho$ number increases in the initial stage
mainly through the process $\rho\leftrightarrow\pi\pi$.  This however
is also the dominant decay mode for the $\rho$ and so the pions are
eventually regained  at freeze-out.
 The three $\rho\pi\pi\gamma$ photon-producing processes discussed in detail
above are not  
enhanced dramatically if the number of $\rho$ mesons increase because 
this is achieved at the expense of reducing the number of pions. As a result,
$\rho\pi\rightarrow\pi\gamma$ in neither favored nor disfavored and
the other two processes $\pi\pi\rightarrow\rho\gamma$ and
$\rho\rightarrow\pi\pi\gamma$ are balanced against each other.
Although all three scenarios are close to the free space result, the 
largest change comes to Scenario~II and the Georgi vector limit, which is 
reduced by almost a factor of two in the final rates as seen in
Fig.~\ref{fig9}.  

%%%%%%%%%%%%%%%%%%%%%%%%%%%%%%%%%%%%%%%%%%%%%%%%%%%%%%%%%%
\vspace{-0.5in}
\begin{figure}
\begin{center}
\hspace{-0.5in}
\epsfxsize=4.0in
\epsffile{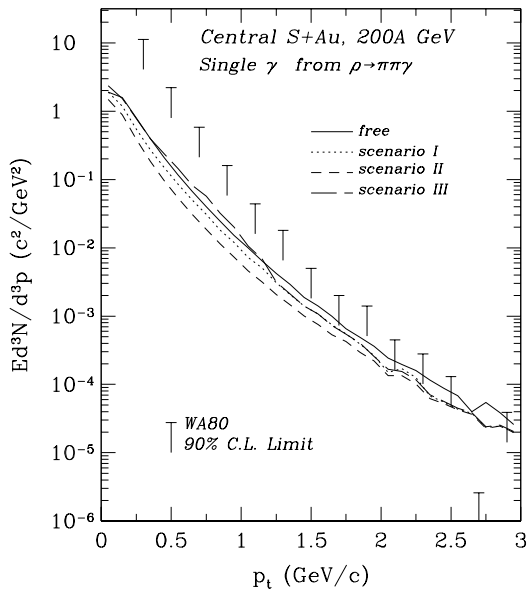}
\vspace{-0.3in}
\end{center}
\caption{\label{fig10} The same as Fig.~\protect\ref{fig9},
for $\rho\to\pi\pi\gamma$.}
\end{figure}
%%%%%%%%%%%%%%%%%%%%%%%%%%%%%%%%%%%%%%%%%%%%%%%%%%%%%%%%%%%%%%

\begin{figure}
\begin{center}
\hspace{-0.5in}
\epsfxsize=4.0in
\epsffile{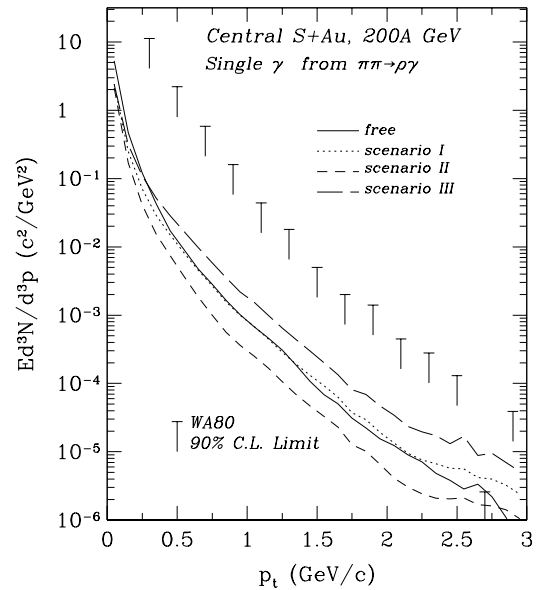}
\vspace{-0.3in}
\end{center}
\caption{\label{fig11} The same as Fig.~\protect\ref{fig9},
for $\pi\pi\to\rho\gamma$.}
\end{figure}

\begin{figure}
\begin{center}
\hspace{-0.5in}
\epsfxsize=4.0in
\epsffile{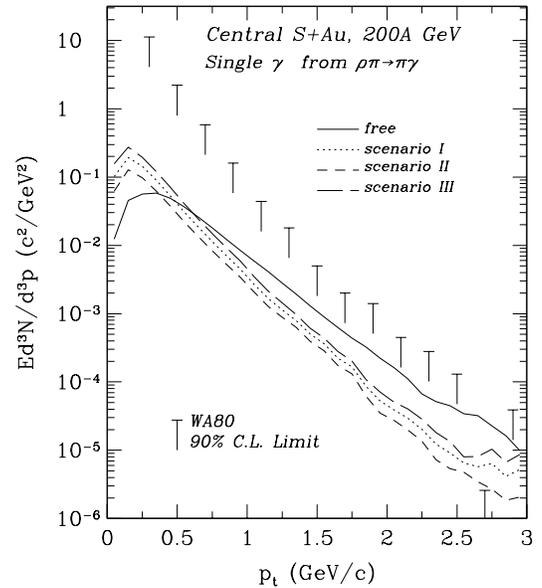}
\vspace{-0.3in}
\end{center}
\caption{\label{fig12} The same as Fig.~\protect\ref{fig9},
for $\pi\rho\to\pi\gamma$.}
\end{figure}

This variation in the three scenarios can be seen better by breaking
the rates up into the important contributions.  This is shown in
Figs.~\ref{fig10}, \ref{fig11}, and \ref{fig12} for the
$\rho\to\pi\pi\gamma$, $\pi\pi\to\rho\gamma$, and $\pi\rho\to\pi\gamma$
rates respectively.  The  main contribution can be seen to come from 
$\rho$ decay, which is larger than  what one would expect from the
thermal rates of Section~III due to off-shell propagation enhancement.
The largest value to the decay rates comes from
Scenario III in which both masses drop, but 
$a$ increases as the ratio of the $\rho$ and $a_1$ mass approaches 1. 
The lowest rates come from Scenario II as one would expect from the
cross-sections.

%%%%%%%%%%%%%%%%%%%%%%%%%%%%%%%%%%%%%%%%%%%%%%%%%%%%%%%%%%
\vspace{-0.5in}
%%%%%%%%%%%%%%%%%%%%%%%%%%%%%%%%%%%%%%%%%%%%%%%%%%%%%%%%%%%%%%
%%%%%%%%%%%%%%%%%%%%%%%%%%%%%%%%%%%%%%%%%%%%%%%%%%%%%%%%%%
\vspace{-0.5in}
%%%%%%%%%%%%%%%%%%%%%%%%%%%%%%%%%%%%%%%%%%%%%%%%%%%%%%%%%%%%%%

It is also interesting to note how similar Scenario I is to the free
rates in all three figures.  This is because Brown-Rho
scaling for the $a_1$ meson does not allow either of the two
parameters of the HLS Lagrangian to change.  The widest range in values
among the scenarios comes in the $\pi\pi\to\rho\gamma$ rates which
give a spread of almost an order of magnitude between Scenario III
(largest) and Scenario II (smallest).  One can see from
Fig.~\ref{fig8} that this process has a larger spread between
scenarios in terms cross sections, too. 

It is interesting that in total, all three scenarios as well as the
free rates touch the upper bounds set by WA80 for large $p_t$.  If
other processes such as $a_1\pi\to\rho\gamma$ were included, they
would also feed into this high momentum region and possibly push the
rates past the upper bounds. One must be cautious however in drawing
conclusions in terms of absolute values of the photon rates at high
$p_t$ even though we cut off the unphysical high-momentum dependence
of our cross sections as discussed at the end of Section~IV. 
Needless to say, results from WA98 and
future photon measurements are needed to properly interpret the discrepancy
with the present data.

With excellent data in the photon spectra, one could possibly
distinguish between the three broad scenarios set forth in this paper.
However, given the extremely poor signal to noise ratio for the
photons, the experimental uncertainty is too large to say whether one
or another of the curves in Fig.~\ref{fig9} is right. One might rank
the scenarios in order of decreasing total photon rate. Then, 
the free-space result and Scenario III seem more likely to exceed 
the upper bounds for the region $p_t<1.75$ GeV, 
 than
Scenarios I and II. Both the free-space result and Scenario III are 
strong in this region for $\rho$ decay.  The free-space result outshines 
all other scenarios for $\rho\pi\to\pi\gamma$, while Scenario III
dominates for $\pi\pi\to\rho\gamma$.
Except for $\rho$ decay above $2$ GeV, Scenario II gives the lowest
predictions. Also, it is the only one to have the same slope
as the WA80 limits.

\section{Summary and Conclusions}

The purpose of this study is to investigate the effect of dropping 
in-medium vector meson masses on photon spectra from heavy ion collisions.
This has been done earlier for dilepton spectra \cite{li95}
and for photon spectra \cite{librown97}. In the case of photon spectra,
the effect of the $a_1$ has to be included \cite{xiong}, since it is
potentially important.

We recalculated three processes important for photon production
in heavy ion collisions using the extended hidden local symmetry 
Lagrangian \cite{hidden} with the proper $a_1$ mass. We find that the
contribution 
of the $a_1$ as intermediate state is relatively small, in contrast to
results found in earlier models \cite{xiong,song93}, but closer to
results in Ref.~ \cite{kim96}. 
It would be interesting to see whether 
 this is indeed a genuine physical effect by looking at alternative
formulations of HLS.  In particular, the $\pi$-$a_1$ mixing is taken
care of now by a shift of the $a_1$ field whereas using one of the
unphysical fields that appear in non-unitary gauges could have the
same effect without modifying the $a_1 \pi \gamma$ vertex. 
Both the $a_1\to\pi\gamma$ decay width and the $s$- to $d$-wave ratio
in $a_1$ decay should be used to identify the most advantageous approach.

However, the small effect of the $a_1$ meson lies in something even
more basic: the linear relation in the HLS between
the coupling $g$ and the vector meson masses.  Therefore, as temperature and
density increase, the masses drop in agreement with Brown-Rho scaling
\cite{br91} and the couplings are driven to zero in accordance with
asymptotic scaling.  This suppresses the rates when density
corrections are taken into account.

Nevertheless, using only the vector and axial-vector
meson masses as input parameters our model reasonably predicts the vector
and axial-vector meson decay widths.  
Our predictions for thermal photon emission rates from
a hadronic gas through the processes 
$\rho\pi\to\pi\gamma$, $\pi\pi\to\rho\gamma$, and $\rho\to\pi\pi\gamma$ are
within the range of similar results in the literature. 
The rates for $\rho\pi\to\pi\gamma$  are slightly higher than those of 
Kapusta, {\em et al.} \cite{kapusta}, but a factor 2-3 lower than those
of Song \cite{song93}.

We considered three different scenarios for the in-medium evolution 
of $m_\rho$ and $m_{a_1}$ in an attempt to understand the
differences in signals they would give: strict Brown-Rho scaling where
the ratio $m_\rho / m_{a_1}$ stays fixed, an
extreme case of the Georgi-vector limit where $m_{a_1}$ is fixed, 
and a schematic model inspired by finite temperature lattice results.

We simulated single photon spectra in central 
S+Au collisions at SPS energies using the relativistic 
transport model that has been used to study dilepton spectra 
in the same reactions.  We included photons from the background
sources of $\pi^0$ and $\eta$ decays, as well as thermal sources
such as meson decays, decays of baryon resonances, and 
two-body processes. We found that more than 95\% of single
photons come from the decays of $\pi^0$ and $\eta$. The 
thermal photons account for only less than 5\% of all single
photons, in agreement with the experimental observation made by
the WA80 collaboration. We compared our thermal photon 
spectra with the experimental upper bound extracted by
the WA80 collaboration. 
Overall, the results were all comparable to each other for the total
rate, none of them exceeding significantly the experimental bounds.
For $p_t<2$ GeV, the largest photon yields were found for the 
lattice-inspired scenario and the simulation without dropping masses.
The Georgi-vector limit case gives the lowest yield in this range.
It slightly exceeds the upper bounds of WA80 for large $p_t$, but a fully
consistent high energy behavior is a matter of further study. 

\bigskip

In conclusion, we find that in the extended hidden local symmetry approach
the role of the $a_1$ for the processes we considered is less important
then it is generally thought. 

We showed that one can implement the dropping in-medium $\rho$ mass and
Brown-Rho scaling without violating the existing experimental limits
for photon production. Furthermore, there are indications that different
scenarios lead to experimentally distinguishable predictions.

\onecolumn

\acknowledgements{
We thank Charles Gale, Mannque Rho and Jac Verbaarschot for stimulating 
discussions and a critical reading of the manuscript and
Madappa Prakash, Ralf Rapp, Edward Shuryak, and Heinz Sorge 
for useful discussions.
This work was supported in part by the US DOE grant DE-FG02-88ER40388
and by the National Science Foundation under Grants No.~PHY-9511923
and PHY-9258270.
}

\appendix
\section{Matrix Elements}

For $\rho^a(p)\to \pi^b(p_1) \pi^c(p_2) \gamma(k)$, the matrix element
${\cal M}$ shown diagrammatically in Fig.~\ref{fig1} is given by the
addition of the following four expressions ($r=m_\rho^2/m_{a_1}^2$).

\ben
{\cal M}_1 &=& \epsilon^{ace} \epsilon^{3be} \frac{e}{2gf_\pi^2}
\left\{m_\rho^2 \epsilon_1 \cdless \epsilon_2 
+ 2r \left( \epsilon_1\cdless \epsilon_2\, p_2\cdless p 
- \epsilon_1\cdless p_2 \, \epsilon_2 \cdless p \right) + 2r^2\left(
\epsilon_1\cdless \epsilon_2 \, p_1\cdless p_2 - \epsilon_1\cdless p_1 \,
\epsilon_2 \cdless p_2 \right) \right\} +
(p_1,b)\leftrightarrow(p_2,c)
\\
{\cal M}_2 &=& \epsilon^{a3e} \epsilon^{bce} \frac{e}{g f_\pi^2}
\left\{ \left( \epsilon_1 \cdless p_1\, \epsilon_2
\cdless p + \epsilon_1 \cdless k\, \epsilon_2 \cdless p_1 - \epsilon_1 \cdless
\epsilon_2\, p_1\cdless k \right) 
\frac{\xi_r(p_1+p_2)^2+m_\rho^2}
{(p_1+p_2)^2-m_\rho^2} 
- \xi_r \epsilon_1 \cdless p_1\, \epsilon_2 \cdless p_2
\right\} + (p_1,b)\leftrightarrow(p_2,c)
\\
{\cal M}_3 &=& 2 age \left( 1+ \xi_r \right) \epsilon^{ace}
\epsilon^{3be} \frac{\epsilon_1\cdless p_2 \,\epsilon_2\cdless
p_1}{(p_1+k)^2-m_\pi^2} + (p_1,b)\leftrightarrow(p_2,c)
\\
{\cal M}_4 &=&  \frac{e}{g f_\pi^2} \epsilon^{ace} \epsilon^{3be}
\left\{ r W \cdless \epsilon_2 + (3r-1) 
\frac{\left( p_1\cdless k\, \epsilon_2\cdless W - \epsilon_2 \cdless p_1\,
k\cdless W \right)}{(p_1+k)^2-m_{a_1}^2}  \right\} +
(p_1,b)\leftrightarrow(p_2,c) 
\een
with $\xi_r=2r(1-2r)$  and 
\ben 
W^\mu = \epsilon_1 \cdless p_2 \left[ (4r-1) p^\mu - rp_2^\mu \right] -
\left[ (4r-1) p_2\cdless p + rp_2\cdless (p_1+k) \right] \epsilon_1^\mu .
\een
As required by gauge-invariance, the total matrix element ${\cal M}$
vanishes under the replacement $\epsilon_2\to k$.

\section{Differential Thermal Rate}

For the process $1+2 \to 3+\gamma$ we define $s=(p_1 + p_2)^2$, 
$t=(p_1 - p_\gamma)^2$. With $E_3=E_1+E_2-E_\gamma$, introduce 
\ben
q_1 = m_1^2 - t;\qquad q_2 = s + t - m_1^2 - m_3^2;
\qquad q_3 = s- m_3^2 = q_1 + q_2;
\qquad
{p'}_k^2 = - m_k^2 + \frac{E_k}{E_\gamma} q_k - q_k^2,\quad k=1,2,3  .
\een
The differential rate is then
\ben
E_\gamma \frac{d R}{d^3 p_\gamma} = \frac{\cal N}{(2 \pi)^7 16E_\gamma^2}
\int\! ds \int\! dt \left| {\cal M}(s,t) \right|^2 \int\! dE_1 \int\! dE_2
\frac{f(E_1) f(E_2)(1+f(E_3))}
{\left[ 4 {p'}_1^2 {p'}_2^2 - \left( {p'}_3^2 - {p'}_1^2 - {p'}_2^2 \right)^2 
\right]^{1/2}}
\een
The integration limits on $s$ and $t$, in addition to $s>(m_1+m_2)^2$, are
such that $q_1>0$ and $q_2>0$. The limits on $E_1$, $E_2$ are set by

\ben
E_1 > \frac{q_1}{4 E_\gamma}+\frac{E m_1^2}{q_1};
\qquad E_2 > \frac{q_2}{4 E_\gamma} +\frac{E m_2^2}{q_2};
\qquad E_1+E_2 > E_\gamma.
\een

\end{document}